\documentclass[9pt,twocolumn,twoside]{osajnl}

\journal{jocn}

\setboolean{shortarticle}{false}
\usepackage{cleveref}
\crefname{equation}{Eq.}{Eq.} 
\crefrangelabelformat{equation}{(#3#1#4)--(#5#2#6)}
\crefname{figure}{Fig.}{Fig.}
\crefname{table}{Table}{Table}

\title{End-to-end Optimization of Optical Communication Systems based on Directly Modulated Lasers}

\author[1,*]{Sergio Hernandez}
\author[2]{Christophe Peucheret}
\author[1]{Francesco Da Ros}
\author[1]{Darko Zibar}

\affil[1]{Department of Electrical and Photonics Engineering, Technical University of Denmark, 2800 Kongens Lyngby, Denmark}
\affil[2]{Univ. Rennes, CNRS, FOTON - UMR6082, 22305 Lannion, France}

\affil[*]{shefe@dtu.dk}

\begin{abstract}
The use of directly modulated lasers (DMLs) is attractive in low-power, cost-constrained short-reach optical links. However, their limited modulation bandwidth can induce waveform distortion, undermining their data throughput. Traditional distortion mitigation techniques have relied mainly on the separate training of transmitter-side pre-distortion and receiver-side equalization. This approach overlooks the potential gains obtained by simultaneous optimization of transmitter (constellation and pulse shaping) and receiver (equalization and symbol demapping). Moreover, in the context of DML operation, the choice of laser-driving configuration parameters such as the bias current and peak-to-peak modulation current has a significant impact on system performance. We propose a novel end-to-end optimization approach for DML systems, incorporating the learning of bias and peak-to-peak modulation current to the optimization of constellation points, pulse shaping and equalization. The simulation of the DML dynamics is based on the use of the laser rate equations at symbol rates between 15 and 25 Gbaud. The resulting output sequences from the rate equations are used to build a differentiable data-driven model, simplifying the calculation of gradients needed for end-to-end optimization. The proposed end-to-end approach is compared to 3 additional benchmark approaches: the uncompensated system without equalization, a receiver-side finite impulse response equalization approach and an end-to-end approach with learnable pulse shape and nonlinear Volterra equalization but fixed bias and peak-to-peak modulation current. The numerical simulations on the four approaches show that the joint optimization of bias, peak-to-peak current, constellation points, pulse shaping and equalization outperforms all other approaches throughout the tested symbol rates. 
\end{abstract}

\setboolean{displaycopyright}{false} 

\begin{document}

\maketitle

\section{Introduction}
Directly modulated lasers (DMLs) are a compelling option for short-reach intensity modulation/direct detection (IM/DD) systems, thanks to their low energy consumption, small form factor and reduced cost \cite{9720189, Huang_2021}. The aim in such systems is to maximize the symbol rate $R_s$ while maintaining sufficient optical received power $P_\mathrm{rec}$ in order to increase net data throughput. This requires the laser to operate in the large-signal regime, where the high modulation index allows high extinction ratio and peak-to-peak output power. However, the large-signal DML dynamics introduce significant waveform distortions, potentially resulting in nonlinear intersymbol interference as the symbol rate is increased \cite{10238466, 10323176}. This is due to the modulation-induced changes in carrier and photon concentration within the laser active region, that cause nonlinear memory effects in the output optical field. Consequently, the DML's response time sets a limit on its modulation bandwidth, restricting the laser's throughput \cite{Coldren2012DiodeCircuits}. 

 \begin{figure*}[ht!]
     \centering
     \includegraphics[width = \linewidth]{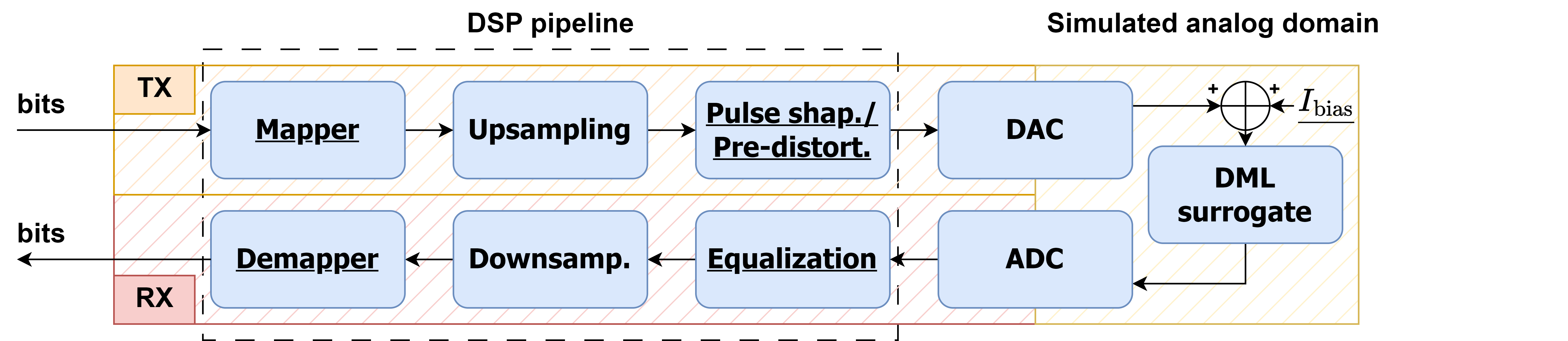}
     \caption{Proposed numerical simulation of a back-to-back DML system. The DSP pipeline at TX (green) and RX (red) aims to encode and decode information subject to the constraints imposed by the simulated analog domain (yellow). The elements optimized in this work are underlined}
     \label{fig:introsetup}
 \end{figure*}
 
Although increasing the bias current to the laser can enhance modulation bandwidth, it comes at the expense of increased energy consumption and lower extinction ratio, resulting in a receiver power penalty. Another alternative to partially overcome bandwidth limitations is tuning transmitter (TX), receiver (RX), digital signal processing (DSP) and the laser-driving configurations (bias and peak-to-peak current to the laser) separately. Yet, the large amount of parameters to be optimized within the DSP pipeline (pulse shaping, pre-distortion, receiver-side equalization) may require a high amount of simulation/system evaluations to converge towards optimal configurations, making such approach intractable in cases where such evaluations are time consuming.

Recent advances in DML development have focused on delivering >100 Gbps bit rates with low energy consumption (<1 pJ/bit) through the use of coupled-cavity laser structures \cite{10080935, 10.1117/12.2656921, 10025553}. The additional optical feedback enabled by such structures enables higher modulation bandwidths through the use of photon-photon resonance (PPR) and detuned loading. 
Impairment compensation in DML-based systems \cite{9720189, Reza:22} has however relied mainly on separate transmitter-side pre-distortion and receiver-side optimization, omitting the potential gains of optimizing both TX and RX jointly. This arises from the fact that the large-signal DML dynamics are governed by nonlinear differential equations, hindering the calculation of analytical gradients and therefore the simultaneous optimization of transmitter and receiver using standard gradient-base optimization techiques. An accurate, differentiable model of the DML dynamics can prove useful in this scenario, allowing the propagation of gradients between TX and RX while accounting for the DML-induced signal impairments. Data-driven modeling is a viable option in this context provided that sufficient data is available, yet the choice of model structure (Volterra filters, neural networks) can pose a challenge. In \cite{10382548}, we conduct a comparison between model structures using the laser rate equations as source of DML waveform data. The results show the potential of transformer-based neural networks \cite{10.5555/3454287.3454758} in the prediction of the DML dynamics.

DSP algorithms have extensively been used to compensate for transmission impairments within optical communication systems. End-to-end (E2E) learning has attracted a special interest in this scope, as it enables the simultaneous optimization of TX (constellation and pulse shaping) and RX, given that a differentiable model of the system under test is available \cite{Srinivasan2023, Minelli23}. Through the use of alternative approaches like gradient-free optimization and reinforcement learning, it is also possible to avoid modeling the communication channel and use local gradient approximations \cite{8792076, Yankov:22}. The focus in E2E learning is to substitute one or several functions of the DSP pipeline at both ends of the channel by an adaptive optimizable function \cite{8054694}. Autoencoders (AEs) are especially popular in this context, as they allow the compression and decompression of data \cite{Goodfellow-et-al-2016} in a similar fashion to how communication systems map symbols to optical waveforms and viceversa \cite{8054694}. Approaches such as geometric constellation shaping (GCS), that have become standard in state-of-the-art in long-haul coherent systems, can be implemented based on this principle \cite{Oliari:22, 10255117, 9852813, 10124361, Yankov:22}. 

In this paper, we demonstrate E2E optimization of TX and RX DSP, together with bias and peak-to-peak modulation current for impairment compensation of DML systems. The proposed simulation approach is shown in \cref{fig:introsetup}. The modeling of the DML dynamics is performed using a data-driven surrogate model, based on the numerical solution of the rate equations as source of data. The waveform data generation uses 4-level pulse amplitude modulation (4PAM) symbols in a back-to-back (B2B) simulation. The proposed AE-based approach is used to optimize DSP configurations at TX (GCS, pulse shaping) and RX (equalization, symbol detection) simultaneously. In addition, the input current offset $I_{\mathrm{bias}}$ and the peak-to-peak modulation current $I_{\mathrm{pp}}$ are used as learnable parameters. 
This provides insight on the optimal compromise between extinction ratio and distortion of the optical waveform. Additionally to the AE approach, three more approaches are included as a benchmark: the uncompensated system without equalization, a receiver-side linear feed-forward equalizer (FFE) and a second E2E approach with learnable pulse shape (LPS) and RX-side Volterra nonlinear equalization (VNLE), but excluding $I_{\mathrm{bias}}$ and $I_{\mathrm{pp}}$ from the optimization. The results show the advantage of the joint optimization of bias, modulation current and DSP for DML systems, with the AE yielding performance gains over the RX-only equalization and a considerable advantage over the VNLE setup in terms of symbol error rate (SER) and mutual information (MI).

This paper is structured as follows: Section~\ref{ch:dmls} describes the dynamic behaviour of DMLs and the state-of-the-art in optical communication laser development. Section~\ref{ch:modeling} analyzes the available algorithms for the modeling of DMLs in terms of complexity, interpretability and compatibility with gradient-based optimization. The fundamentals of E2E learning and its application to DMLs in the literature is reviewed in Section~\ref{ch:e2e}. The motivation and structure of the proposed DML model and the optimization approach around it can be found in Section~\ref{ch:setup}. The results of the E2E-optimized B2B DML system simulation are described in Section~\ref{ch:results}. The conclusions are summarized in Section~\ref{ch:conclusion}.

\section{DML dynamics} \label{ch:dmls}
The rate equations governing the photon density S(t), carrier density N(t) and phase $\phi(t)$ in DMLs are given by \cite{Coldren2012DiodeCircuits}:

\begin{align}
    &\frac{dS(t)}{dt} = \Gamma g_0 (N(t) - N_0) \frac{1}{1 + \epsilon S(t)} S(t) - \frac{S(t)}{\tau_p} + \frac{\Gamma \beta N(t)}{\tau_n} \label{eq:phos} \, , \\
    &\frac{dN(t)}{dt} = \frac{I(t)}{qV} - \frac{N(t)}{\tau_n} -g_0(N(t)-N_0)\frac{1}{1+\epsilon S(t)}S(t) \label{eq:carrs} \, ,\\
    &\frac{d\phi(t)}{dt} = \frac{1}{2} \alpha \left[ \Gamma g_0 (N(t) - N_0) - \frac{1}{\tau_p} \right] \, , \label{eq:phase}
\end{align}

where $\Gamma$ is the mode confinement factor, $g_0$ is the gain slope constant, $N_0$ is the carrier density at transparency, $\epsilon$ is the gain compression factor, $\tau_p$ is the photon lifetime, $\beta$ is the fraction of spontaneous emission coupled into the lasing mode, $\tau_n$ is the electron lifetime, $I(t)$ is the injected current, $q$ is the electron charge, $V$ is the active layer volume and $\alpha$ is the linewidth enhancement factor. The output optical power is given by:

\begin{equation}
    P(t) = \frac{S(t) V \eta_0 h \nu }{2 \Gamma \tau_p} \, ,
    \label{eq:pout}
\end{equation} 

where $\eta_0$ is the differential quantum efficiency, $h$ is Planck's constant, and $\nu$ is the unmodulated optical frequency. Thus, the modulated optical power $P(t)$ is proportional to the photon density $S(t)$, while the rate of change of the optical phase (instantaneous angular frequency) is proportional to the carrier density over threshold.

\subsection{Small-signal regime}
The term $N(t) - N_0$ in \cref{eq:phos,eq:carrs} hints the reservoir-like behaviour of the carrier density fluctuation in the laser active region. When $I$ reaches the laser's threshold current level $I_{\mathrm{th}}$ (optical gain overcomes cavity losses) the carrier density reaches its maximum steady-state level $N_0$ (the reservoir becomes full). The carrier excess generated by $I - I_{th}$ "overflows" the reservoir, generating net stimulated emission of photons. The higher $I-I_{th}$ is, the more photons are emitted, and through \cref{eq:pout} the larger $P(t)$ will be. This also entails that the stimulated recombination of carriers increases, bringing $N(t)$ back to its threshold level $N_0$ after some transient period. The interaction between carriers and photons creates a damped oscillatory behaviour in $S(t)$ and $N(t)$ (carrier-photon resonance) when $I(t)$ is modulated, as the increase of one drives the decrease of the other until steady state is reached. The optical gain generated by the laser, dependent of $N(t)$ and $S(t)$, is given by \cref{eq:gain}:

\begin{equation}
    g(N,S) = \frac{g_0}{1+\epsilon S(t)} \ln\left (\frac{N(t)+N_s}{N_0 + N_s}\right) \, ,
    \label{eq:gain}
\end{equation}

where $N_s$ is a fitting parameter used to ensure the logarithm is finite and defined for $N(t) > 0$. Both $g_0$ and $N_s$ are usually fitted from the measured response of the laser \cite{Yamaoka2021}. Given that the optical gain $g$ is monotonically (although non-linearly) related to $N$ and $S$, it can be assumed that for small variations of the carrier density $\Delta N$ the optical gain variation is proportional to $\Delta N, \Delta S$. This leads to $g = g_{th} + a\Delta N - a_p\Delta S$, using the local slopes $a = \partial g/\partial N$ (differential gain) and $a_p = \partial g/ \partial S$. The approximation yields accurate results as long as $I_{\mathrm{pp}} << I_{\mathrm{bias}}$ for $I > I_{th}$, the small-signal regime conditions. Although such conditions are usually impractical in a communication setting, the small-signal analysis provides valuable insight into the characteristics of a DML. It can be shown that, under the small signal approximation and under sinusoidal excitation at angular frequency $\omega$, the frequency response of the laser can be expressed according to:

\begin{equation}
    H(\omega) = \frac{\omega_R^2}{\omega_R^2 - \omega^2 + j\omega \gamma} \, ,
    \label{eq:s21}
\end{equation}

where $f_R$ is the natural resonant frequency of the system, $j$ is the imaginary unit and $\gamma$ is the laser's damping factor. Assuming operation over threshold, $f_R$ can be approximated as \cite{Yamaoka2021}:

\begin{equation}
     f_R \approx \frac{1}{2 \pi}\sqrt{\frac{v_g a \bar{S}}{\tau_p}}  = \frac{1}{2 \pi} \sqrt{\frac{\Gamma v_g a}{qV} \eta_i (I_{\mathrm{bias}} - I_{th})} \, , 
     \label{eq:reloscsimpl}
 \end{equation}
 
where $\bar{S}$ denotes the average photon density. $f_R$ is also called relaxation-oscillation or carrier-photon resonance frequency, as it determines the frequency of the transient damped oscillations created in $S(t)$ and $N(t)$ due to the aforementioned carrier-photon interaction under current modulation. $\gamma$ can be expressed as:

 \begin{equation}
    \gamma = K f_R^2 + \gamma_0 \, , 
\end{equation}

where $K$ is the "K-factor" governing the laser response at high modulation frequencies and $\gamma_0$ is the damping offset. Both parameters are usually obtained through fitting from the modulation response curve (\cref{eq:s21}). 

\begin{figure}[t]
    \centering
    \includegraphics[width = \linewidth]{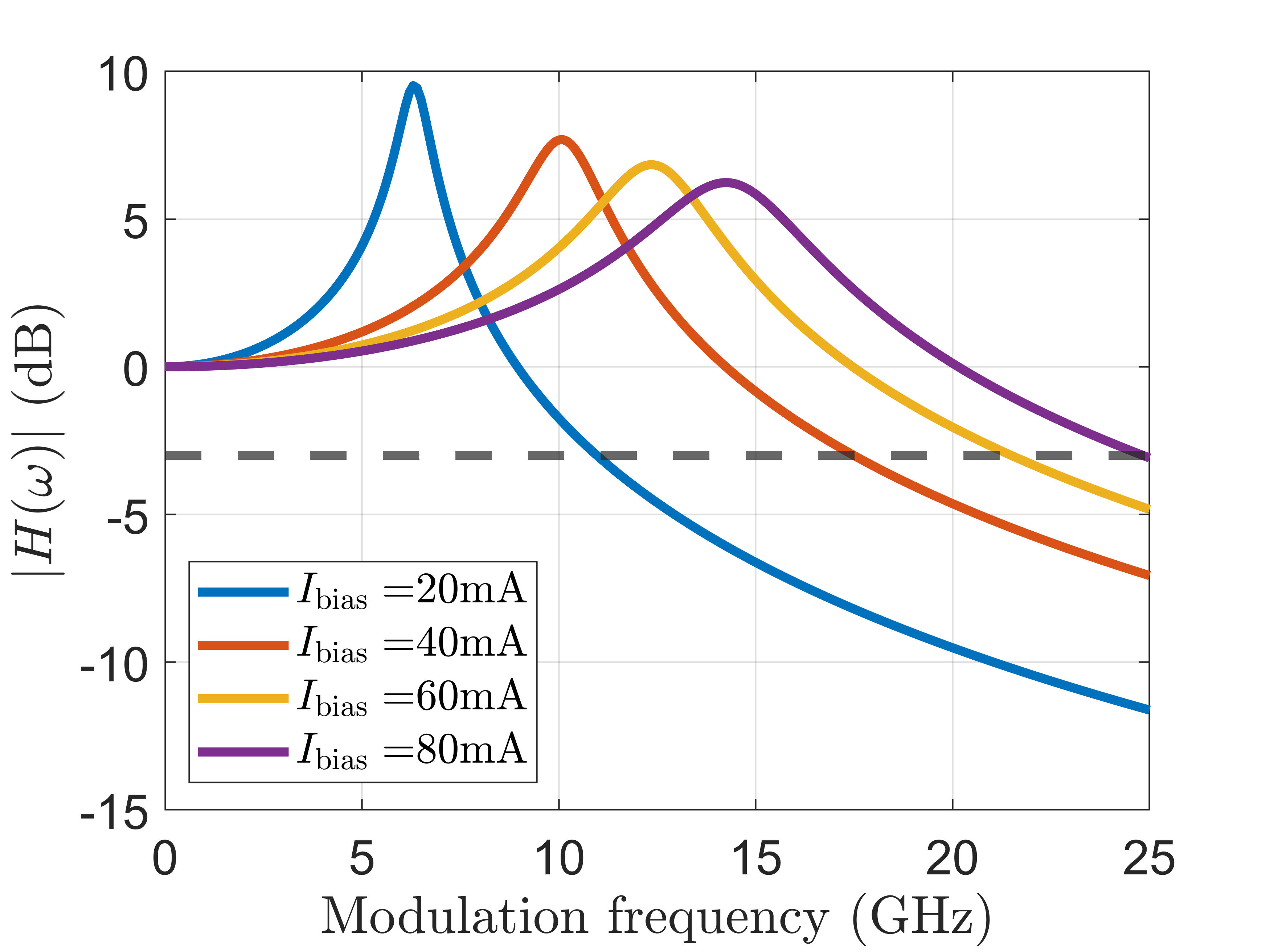}
    \caption{Small-signal modulation response of a simulated DML. The dotted line shows the 3 dB bandwidth level}
    \label{fig:s21}
\end{figure}

The linear relation between $\omega_R^2$ and $(I_{\mathrm{bias}}-I_{th})$ describes the impact of $I_{\mathrm{bias}}$ in the modulation response of the laser. Such relation is shown in \cref{fig:s21}, where various frequency responses $|H(\omega)|$ are represented for different values of $I_{\mathrm{bias}}$. It becomes apparent how the response shows a peak (that coincides approximately with $f_R$) and a rapid damping for frequencies beyond it. A 3-dB bandwidth $f_{3dB}$ can be defined as the frequency for which the magnitude of the modulation response decreases to half of its DC value. For small $\gamma$, the approximation $f_{3dB} \approx 1.55 f_R$ is often used. The small-signal intensity modulation (IM) of DMLs is therefore dictated by the $I_{\mathrm{bias}}$ level, with higher levels associated with a higher 3dB-bandwidth, although with diminishing returns due to the $(I_{\mathrm{bias}}-I_{th})^{1/2}$ factor in \cref{eq:reloscsimpl}.

\subsection{Large-signal regime}
The cost and power constraints in short-reach IM/DD systems favor amplifier-free operation. The large-signal regime describes the DML laser behaviour in most communication settings, where a high modulation index is desirable to overcome receiver noise while avoiding the use of amplifiers. This is due to the analytical solutions to the rate equations being unavailable in this regime (where $I_{\mathrm{pp}}$ and $I_{\mathrm{bias}}$ are comparable in magnitude). The most direct effect of a large $I_{\mathrm{pp}}$ is the introduction of signal intensity distortion due to the effect of relaxation oscillations. This effect is depicted in \cref{fig:lsdiags}a, where overshoot and undershoot can be observed right after changes of the modulating current value. The larger instantaneous variation of $N, S$ enhances the amplitude of the relaxation oscillations, introducing signal components uncorrelated to $I$. The instantaneous emission frequency $\nu$ can also show significant differences with respect to the small signal regime, as shown in \cref{fig:lsdiags}b. The figure shows the variation of $\nu$ resulting from a current modulation. This is due to the signal chirp and relaxation oscillations resulting from the higher modulation index. Frequency chirping can induce a critical limitation of the transmission distance over a dispersive optical fiber. This is due to the interaction between chirp and CD in optical fiber, that induces pulse broadening on the envelope of the received optical signal. The effect of chromatic dispersion on a signal with nonlinear chirp is depicted in \cref{fig:lsdiags}c, d, where the eye diagram of the simulated optical signal before and after transmission over 2 km of standard single-mode fiber (SSMF) is shown. The detrimental impact of pulse broadening is twofold: it decreases the peak optical power of the pulses while introducing ISI between neighboring symbols. It therefore entails substantial worsening of the effective SNR on top of the linear fiber-induced attenuation. The instantaneous chirp expression \cite{Bowers84, 4271197} is obtained from the derivative of \cref{eq:phase}:

\begin{figure}[t]
    \centering
    \includegraphics[width = \linewidth]{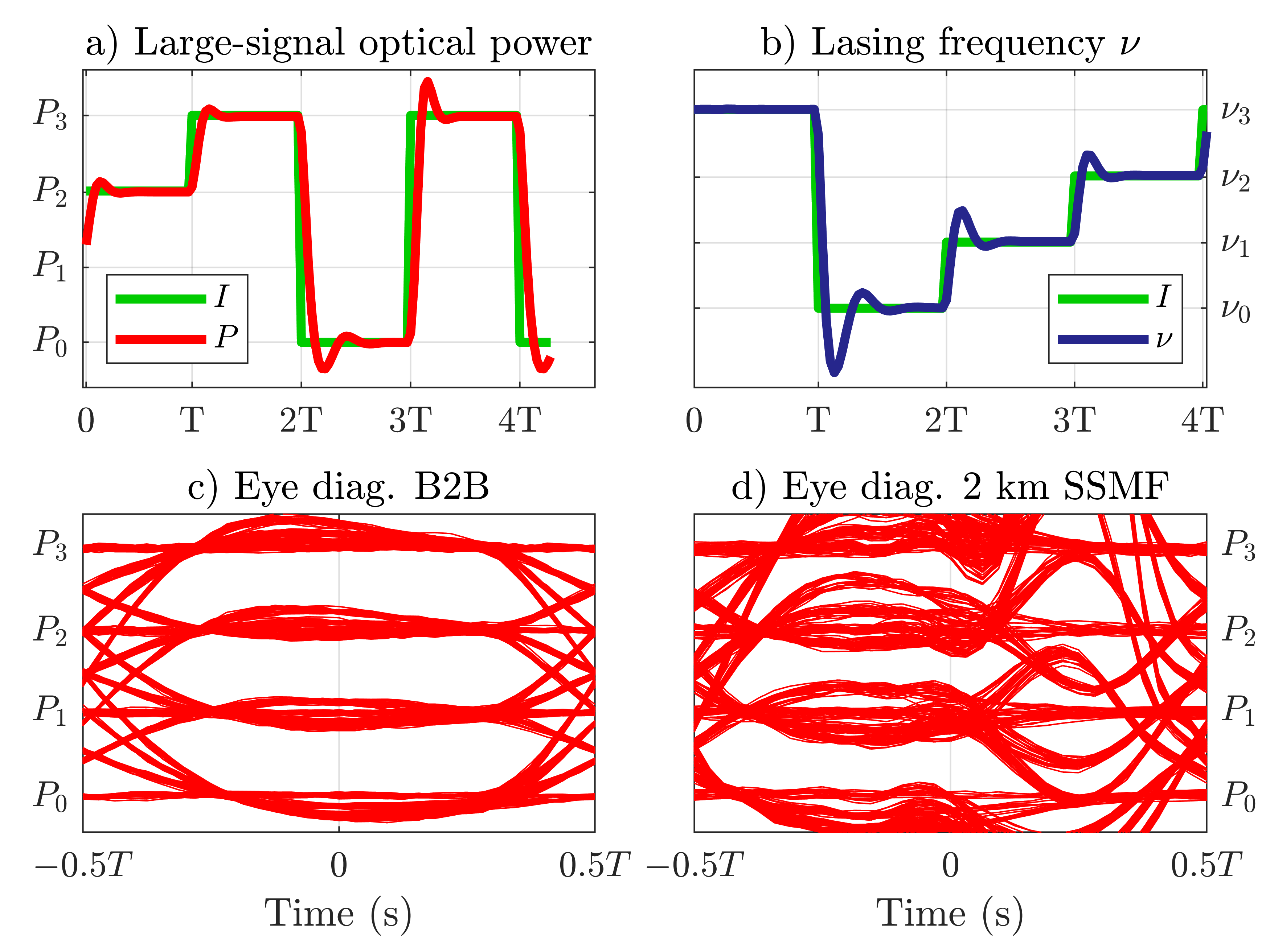}
    \caption{Waveform distortion introduced by DMLs: a) relaxation oscillations, b) instantaneous chirp, c) simulated back-to-back eye diagram under 4PAM modulation and d) simulated eye diagram after 2km of SSMF, $R_s = 30$ Gbaud, central wavelength $\lambda = 1500$ nm. The power levels $P_{0}, P_{1}, P_{2}, P_{3}$ and instantaneous lasing frequency levels $\nu_{0}, \nu_{1}, \nu_{2}, \nu_{3}$ correspond to the 4 symbols in the 4PAM intensity modulation of $I$.}
    \label{fig:lsdiags}
\end{figure}

\begin{equation}
    \Delta \nu = \frac{\alpha}{4\pi} \left[\frac{1}{P(t)} \frac{dP(t)}{dt} + \kappa P(t) \, \right] ,
    \label{eq:chirp2}
\end{equation}

and the factor $\kappa$ is defined by:

\begin{equation}
    \kappa = \frac{2 \Gamma}{\eta h \nu_0 V} \epsilon \, ,
\end{equation}

where $h$ is Planck's constant. The first factor in the sum of \cref{eq:chirp2} is known as transient chirp, while the second is called adiabatic chirp. The prevalence of each chirp factor depends on the dynamics of $P$, and therefore on the variation of the input current $I$. The derivative term of the transient chirp makes it sensitive to fast variations of the optical power. This makes it especially prevalent in high $R_s$ conditions, where the rapid variation of power makes its derivative large. The use of sharp pulses, like square pulses, induce high instantaneous derivative and high relaxation oscillation amplitude, leading to additional transient chirp in the output waveform. The adiabatic chirp becomes prevalent under high $I_{\mathrm{bias}}$ and optical output powers, especially at lower $R_s$. Chirp in DMLs is therefore unavoidable, but the careful selection of certain configuration parameters ($I_{\mathrm{pp}}, I_{\mathrm{bias}}$, pulse shape) could help mitigate it. 

The limited modulation bandwidth of DMLs is associated with several factors, including the aforementioned carrier-photon interaction. Such factors introduce linear and nonlinear memory effects, that become apparent as $R_s$ increases. Although \cref{eq:s21} gives meaningful insight on the characteristics of the DML, the DML response depends on the laser configuration, including $I_{\mathrm{bias}}, I_{\mathrm{pp}}$, and pulse shaping, and can only be obtained through numerical simulations. Additionally, the susceptibility of the system to timing and amplitude impairments varies depending on the channel characteristics. Thus, the use of eye diagrams (\cref{fig:lsdiags}c and d) is widely used in the evaluation of experimental DML-based systems in the large signal regime, as it allows to qualitatively assess the signal degradation at the receiver.

\begin{figure}[t]
    \centering
    \includegraphics[width = \linewidth]{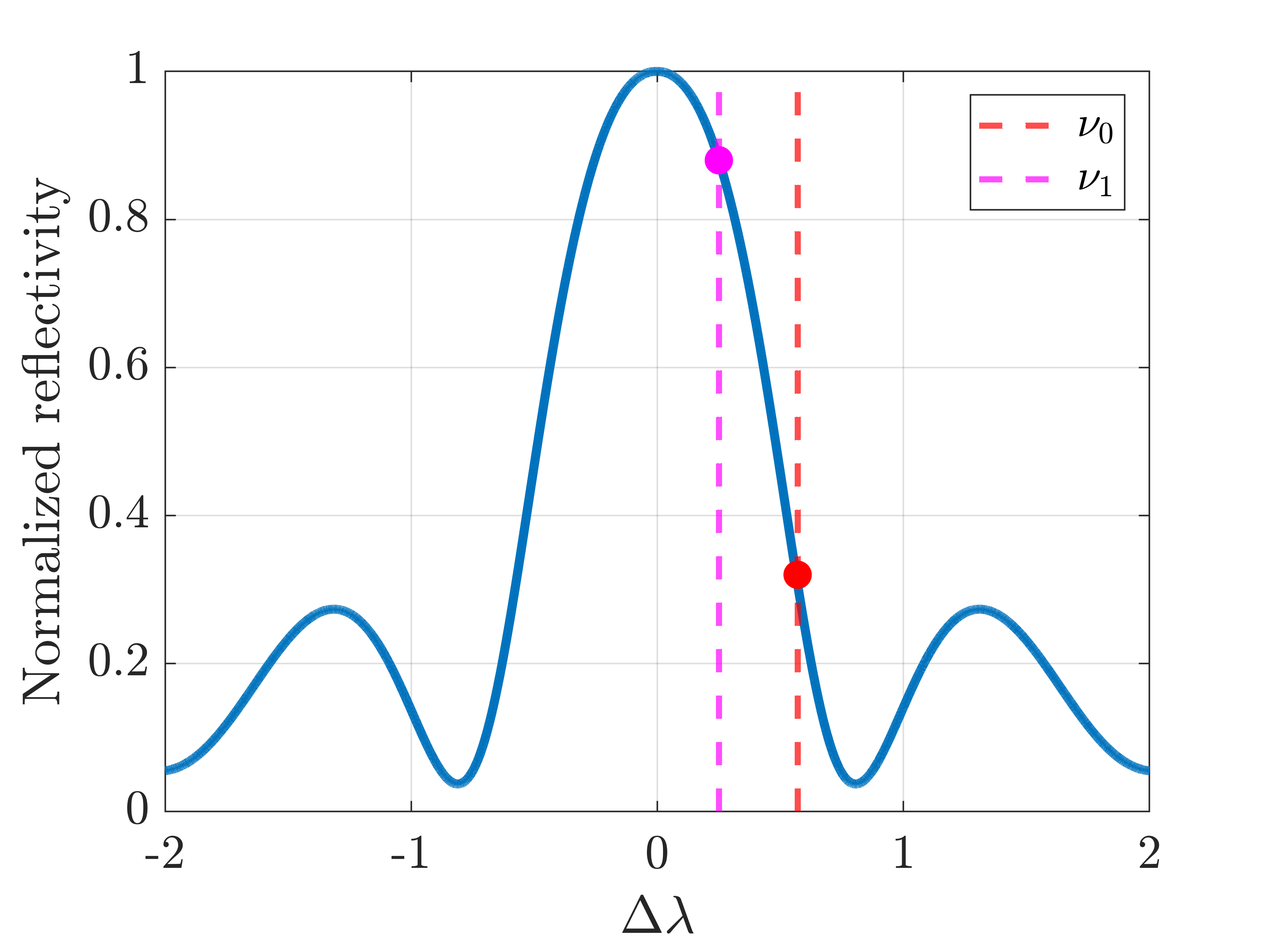}
    \caption{Reflectivity vs detuning of a DBR. The modulation-induced chirp drives the lasing wavelength towards the maximum reflectivity peak}
    \label{fig:detload}
\end{figure}

\subsection{Coupled-cavity effects}
Through the use of additional passive and active laser sections in the laser structure, it is possible to overcome the physical limitations that hinder the DML modulation bandwidth beyond 40 GHz \cite{8332470, 9345388, 7811302, 9288898, Yamaoka2021}. Among the various techniques used, detuned loading and photon-photon resonance (PPR) have delivered promising results in several implementations. Detuned loading is based on the combined effect of distributed reflectors, mostly distributed Bragg reflectors (DBRs) and modulation-induced chirping in order to enhance the dynamic response of the DML \cite{736110, 5518532}. This is achieved through the design of the gain and reflective sections, so the range of main lasing wavelengths is detuned from the main lobe of the DBR reflectivity spectrum. \cref{fig:detload} shows the instantaneous lasing frequencies $\nu_0, \nu_1$, associated with 2 different pump current levels $I_0, I_1$. Both frequencies fall on significantly different regions of the DBR reflectivity spectrum, leading to a difference in mirror loss and the effective differential gain $a_{eff}$. The approximation of $f_R$ in \cref{eq:reloscsimpl} must therefore be modified to account for the detuning $\Delta \lambda$ from the DBR central wavelength:

\begin{equation}
    f_{R, DL} \approx \frac{1}{2 \pi}\sqrt{\frac{v_g S}{\tau_p} \Re\left[\Tilde{\Gamma}_z(\Delta \lambda)g_0(1+j\alpha)\right]} \, ,
\end{equation}

where $\Tilde{\Gamma}_z(\Delta \lambda)$ is the complex reactive confinement factor in the active section \cite{5518532}. The larger the imaginary part of this factor becomes, the larger the impact of $\alpha$ (proportional to chirp) becomes, enhancing $f_R$. When $\Delta \lambda = 0$, $\Tilde{\Gamma}_z(\Delta \lambda)$ becomes real, and therefore no $f_R$ advantage is obtained from the passive section. Detuned loading is therefore an effective method to exploit the FM derived from direct modulation by enhancing the IM, resulting in higher modulation bandwidth. This is not only desirable from the IM bandwidth perspective, but it can also mitigate chirp, leading to reduced distortion after fiber transmission \cite{Matsui2021}.

PPR bandwidth enhancement is based on the interaction between longitudinal modes in the laser cavity. When such modes are close in wavelength, they can still be coupled, \cite{6472009, 892568} leading to a potentially faster modulation response compared to the single-mode configuration. This is usually done through amplification of a secondary side mode using a coupled reflective cavity, as in the case of detuned loading. Both approaches are mutually compatible, resulting in further enhancements in modulation bandwidth. In the case of PPR, a key design goal is to tune the grating phase to match the round-trip phase of the main mode and the side mode, forcing constructive interference between themselves. In this fashion, $\Gamma$ shows also a time dependency, due to the time-varying interaction between the modes in the cavity. The combined gain of the two modes makes possible to force a second resonance peak, of higher frequency that the carrier-photon resonance, to appear in the modulation response of the DML.

\subsection{High-speed DMLs}
The modulation performance of DMLs, based on both in-plane lasers \cite{Huang_2021, 9345388, 9874980, Shinohara:23, Ohno:23, 10041217} and vertical cavity surface-emitting lasers (VCSELs) \cite{photonics9020107, Hoser:22, Wu:22, 9979298, Ge:23} structures, has been extensively developed through cavity design. It must be noted that the commercial interest of DMLs often resides in their energy efficiency and cost, and considerations like reliability, output power and thermal performance play a role in their development. Their comparison must therefore go beyond modulation bandwidth, and a more holistic assessment must be made \cite{Huang_2021}.
Within the cavity design choice, several structures and substrates have been employed in the development of high-bandwidth ($f_{3dB}$) DML lasers. >50 GHz bandwidth was achieved in \cite{10241987} using push-pull modulation, where the current is injected on two locations within the cavity. The driving of the laser is designed so an increase in one of the currents leads to a decrease in the other, and viceversa. Through the combined use of a MQW active region and a single DBR section enhancing detuned loading and PPR, uncooled >100 GHz bandwidth was obtained in \cite{10025553}, maintaining >70GHz at 85°C. The combination of detuned loading and PPR has been used in several works, obtaining $f_{3dB}$ between 40 and 65GHz \cite{10080935, 10.1117/12.2656921, 9720189, Che:20}. As a general overview, several configurations have achieved bandwidths over 50GHz, but 100 GHz have been reached using more complex and costly fabrication processes \cite{Huang_2021}.

\section{System Modeling} \label{ch:modeling}
Modeling a dynamic system is a fundamental first step to its optimization. In the case of DMLs, several alternatives are available depending on the computational resources available, time constraints and desired accuracy and exhaustiveness of the modeling. In this section we introduce some of the most common approaches, sorting them from the more physics-intensive ones to the purely data-driven. 

\begin{figure}[t!]
    \centering
    \includegraphics[width=\linewidth]{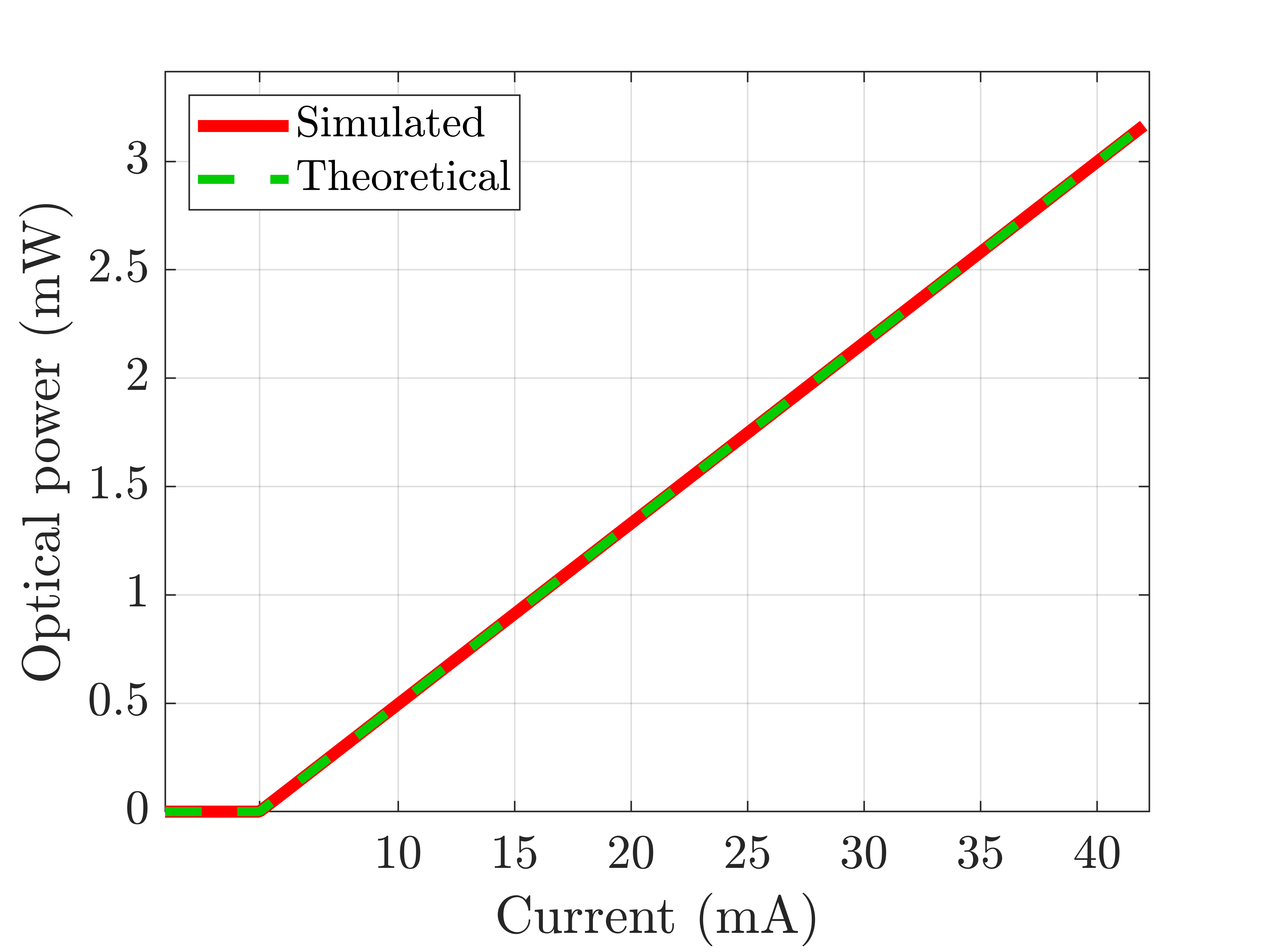}
    \caption{P-I curve associated with the simulated DML. Depending on the physical structure of the laser and thermal effects, the power response of the laser might show nonlinearities}
    \label{fig:picurve}
\end{figure}

\subsection{Parameter extraction and ODE solvers}
Given that the previously introduced rate equations (\cref{eq:carrs,eq:phos,eq:phase}) yield accurate prediction of the DML response in line with experimental verification \cite{9872530, 495166, 571646}, they may seem as the most direct way of estimating the system response. However, they entail 3 main challenges:

\begin{itemize}
    \item Most of their parameters are not easily measurable
    \item They require the use of numerical ordinary differential equation (ODE) solvers in the large signal regime
    \item Analytical gradients are unavailable due to the computational structure of ODE solvers
\end{itemize}

The problem of parameter extraction has been extensively treated in the literature \cite{9872530, 495166, 571646, TOMKOS2001109}. Iterative techniques allow to obtain all the relevant parameters based on experimentally obtained data, like the lasing spectrum, the modulation response (\cref{fig:s21}), the static light-current (L-I) curve (\cref{fig:picurve}) or the spontaneous emission spectrum of the laser. Through the use of machine learning, it is possible to automatize this process, simply by providing the algorithm with the necessary figures of merit, avoiding manual calculation of parameters \cite{Marchisio:24}. Yet, quantities like $N_0$ are not directly measurable, and must be estimated through related parameters, leading to potential inaccuracy.

Even when all the necessary parameters are available, the use of ODE solvers may prove impractical in some scenarios. The large values of $S$ and $N$ can lead to instability in the numerical calculation, leading to divergence and failure of the method. This makes convergence analysis a must have when using ODE solvers, and re-initialization of the calculation may be needed \cite{10.1002/mop.22180}. The sequential nature of the widely used Euler-based methods \cite{Butcher20161} makes parallelization challenging, leading to computational bottlenecks depending on the analyzed system. The largest advantage of such methods is their configurable precision, but the computational toll associated with high precision must also be considered. Lastly, ODE solvers make use of local gradient approximations in order to solve differential equations. This collides with the concept of automatic differentiation that many machine learning and optimization frameworks use, where systems are usually built on functions with known analytical derivatives \cite{Griewank_2003, 10.5555/3122009.3242010}, making gradient calculations faster. This drawback makes an automatically differentiable model of DMLs desirable, as it may allow the simultaneous optimization of large parameter spaces in a relatively reduced computation time. The next subsections will describe some of the available alternatives for the modeling of DMLs. 

\subsection{Circuit-equivalent models}
Circuit-equivalent models \cite{10.1002/mop.22180, Feng2004, Ding:23, 1070628, Habermayer1981, tucker81, 1250490} represent a less complex alternative to rate equation solvers, while sharing the interpretability of physics-based models. They make use of electrical components (amplifiers, capacitors, resistors...) to model the dynamic behaviour of the laser by relating their characteristics to the laser's rate equation parameters. Thus, they are able to reproduce the behaviour of a wide variety of lasers, as long as the corresponding parameters can be measured. The use of current and voltage signals as sole dynamic variables makes them an ideal candidate for circuit simulators, thus facilitating the task of building a numerical simulation. Some literature has included the Langevin phase and intensity noise sources in the simulation, \cite{Horri2012, 1035984} thus introducing significant acceleration with respect to noisy rate equations, where stochastic local gradients can worsen the performance of adaptive step solvers. 

On the other hand, circuit-based models share some of the pitfalls of ODE solvers. Firstly, although circuit simulators are optimized for the task, exhaustive models can be relatively complex, introducing significant computational burden to the simulation \cite{1250490}. This becomes especially true in optimization scenarios, where circuit-specialized software cannot be utilized, and the numerous differential relations between variables must be obtained numerically, i.e. with ODE solvers. In conclusion, although they can be useful in the design and simulation of DMLs, circuit-equivalent models are so far impractical as part of a gradient-based optimization pipeline, and automatically differentiable alternatives should be developed for this purpose \cite{Srinivasan2023}. 

\begin{figure}[t]
    \centering
    \includegraphics[width = \linewidth]{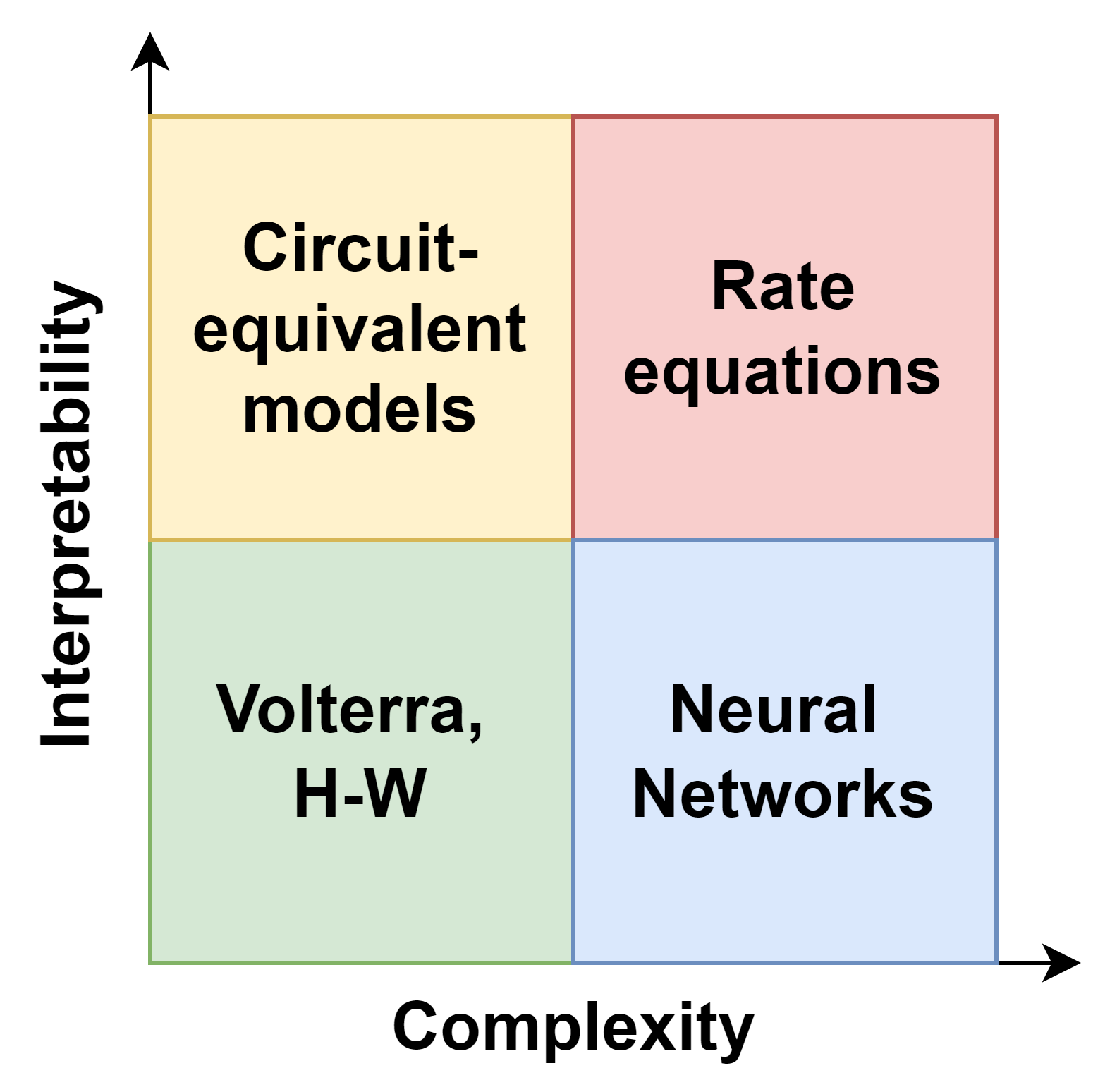}
    \caption{Relation between complexity and interpretability of the proposed modeling techniques}
    \label{fig:complxphys}
\end{figure}

\subsection{Interpretable data-driven modeling}
The mathematical modeling of dynamic systems (based on underlying physics, data, or a mixture of both) has been studied extensively in the control theory community, in what is usually called system identification \cite{systemidt, nelles}. A fundamental advantage of data-driven modeling is its compatibility with numerical optimization techniques: they are mostly based on continuous, easily differentiable functions that provide numerical stability in the search of optimal configurations. Some of the most popular approaches for discrete linear systems are variations of autoregressive such as integrated moving average models (ARIMA) \cite{5953012, 8350823, 10393251}. These models combine the use of 3 techniques: the evaluation of past outputs to predict future ones (AR), the discrete differentiation of the time series in order to force its stationarity (I) and a moving average (MA) of the past prediction errors. However, due to the nonlinear nature of the DML behaviour, many classical system identification models cannot be applied. Additionally, ARIMA models are based on endogenous time series, i.e. the sequence they make predictions on is determined solely by its past values. In order to model nonlinear dynamics, higher-order temporal dependencies and/or nonlinear activations are often used. Nonlinear autoregressive exogenous models (NARX) \cite{LJUNG20201175, 8897147} are a common choice in this case, as they combine linear operations with a static nonlinearity that allows to model a larger function space. Defining an input $u(n)$, an estimated output $\hat{y}(n)$, a input feature vector $x(n)$, a learnable parameter space $\theta$, a and a static nonlinear function $h$, a generic NARX model expression can be obtained:

\begin{equation}
     x(n) = [\hat{y}(n-1), ... \hat{y}(n-N_y), u(n), ...,  u(n-N_u)] \, ,\\
\end{equation}
\begin{equation}
        \hat{y}(n,\theta) = h(x(n),\theta) \, ,
\end{equation}

where $N_u, N_y$ are the input and output memory length, respectively. This general definition includes a large variety of functions, with different nonlinearities, recursivity schemes and complexity levels. 

Volterra filters are often used in communication-oriented DSP to model and/or compensate for the dynamics of nonlinear system \cite{Chen:17, 9780595, 8037970, 7110510}. They are among the simplest NARX models, as they do not employ recursion, and nonlinearity is achieved through multiplication of past input samples with each other. They are based on Volterra series \cite{ELLIOTT2001367}, a variation of the Taylor series where instead of evaluating the analyzed function around a single point, it is evaluated over an infinite range of past samples. Given the previous definitions of $\hat{y}(n), u(n)$ the discrete-time Volterra series can be evaluated as:

\begin{equation}
\begin{aligned}
    \hat{y}(n) = h_0 + \sum_{k_1 = 0}^{\infty} h_1(k_1) u(n-k_1) + \\ \sum_{k_1 = 0}^{\infty} \sum_{k_2 = 0}^{\infty} h_2(k_1, k_2) u(n-k_1) u(n-k_2) + ...
    \label{eq:voltser}
\end{aligned}
\end{equation}

where $h_0, h_1, ..., h_n$ are the Volterra kernels and $k_1, k_2, ..., k_n$ are the delays associated with every kernel order $1, 2, ... n$. As in the case of the Taylor series, increasing the order of the series delivers potentially higher accuracy in the representation, at the cost of increasing complexity \cite{10.1049/el_19960544, 534897}. This forces real implementations to truncate the Volterra series in both kernel order and delay, leading to Volterra filters. It must be noted that the Volterra filter of order 1 corresponds to the linear convolution operator, and therefore the Volterra kernels can be interpreted as a higher-order impulse response of the analyzed system. \cref{fig:complxphys} shows a comparison in terms of interpretability and potential complexity between data-driven and physics-driven modeling methods. The main challenge when using Volterra filters is the calculation of optimal kernels. There are several techniques developed for this purpose \cite{Korenberg1988, eda737, Orcioni2005}, but numerical multivariate optimization techniques can also be used at the expense of higher computation time. 

Hammerstein-Wiener (H-W) models \cite{LJUNG20201175} are an alternative to NARX-based system identification, and provide design flexibility while including a larger number of hyperparameters to be tuned. They are based on the combination of an input static nonlinear function $h$, a linear infinite impulse response (IRR) filter $z(n)$ defined by the real or complex coefficients $p_i, q_i$ and an output static nonlinear function $g$. Its mathematical expression is:

\begin{equation}
    w(n) = h\left[u(n)\right] \, , 
\end{equation}
\begin{equation}
    z(n) =  \sum_{i=1}^{N_z} q_i z(n-i) + \sum_{j=0}^{N_u} p_i w(m-j) \, , 
\end{equation}

\begin{equation}
    \hat{y}(n) = g \left[ z(w(n)) \right] \, , 
\end{equation}

where $N_w, N_z$ are the number of input and output delays, respectively. The choice of $h, g$ and the delay orders $N_w, N_z$ must be tailored to the specific system to be modeled. Although the nonlinearity choice is usually hyperparameter-optimized among a pool of usual functions (sigmoidal, piece-wise continuous, wavelet), the linear transfer functions can be estimated through gray-box modeling (combination of a-priori physical knowledge with data-driven approaches), numerical optimization or stochastic models like the expectation-maximization (EM) algorithm can also be used \cite{WILLS201370}. One of the main advantages of H-W models resides in their interpretability, given that the linear transfer function governing them can be analyzed with traditional spectral analysis as any other linear filter (Laplace/Z transforms, etc). When the physics of the nonlinear system to be modeled are unknown or too complex to be represented in such structure, purely data-based approaches, like neural networks (NNs), are a popular approach \cite{nelles}. 

\subsection{Neural networks} 
NNs are mathematical structures where complex calculations would be performed through the combination of large numbers of simpler operations \cite{Goodfellow-et-al-2016}. Their potential resides in the use of relatively simple linear functions followed by a nonlinear activation. Although, as in the case of Volterra filters, they can be considered a NARX model, their advantage resides in their structural flexibility, that allows them to specialize depending on the task to be performed. The concatenation of simple mathematical operations within NNs leads to large, usually non-orthogonal parameter spaces. Such spaces must be optimized to meet a certain objective, abstracted into a loss or cost function. This is usually done through gradient-based numerical optimization techniques, where instead of calculating the overall function gradient analytically, it is approximated based on reduced subsets of the training data (mini-batches). This procedure accelerates the overall gradient calculation, leading to high performance models without explicit physical knowledge of the system under investigation. The main inconvenient of this procedure is however their lack of interpretability: the concatenation of nonlinearities within NNs them makes it difficult to understand the interaction between the different parameters in the network. 

NNs have been widely used in the modeling of optical communication systems and subsystems \cite{9802856, Song:23, 9905906, 9090286}. Many of the NN architectures in the field have tried to embed some temporal context into the network, as many of the systems found in communications have memory elements. Time-delay neural networks (TDNNs) \cite{WAIBEL1990393, 8401897} are a special case of feedforward neural networks (FFNNs), where no recursive element is added to network. Instead, the input of the network is built as a sliding window of past samples, thus providing the network with temporal context while maintaining relatively low complexity. Single-dimensional convolutional neural networks (CNNs) \cite{10.5555/303568.303704} work on the same principle as TDNNs, although they sometimes inherit pooling layers from its higher-dimensional counterparts. Another possible approach to address this problem is using recurrent neural networks (RNNs), where the network output depends not only on the input features, but also on the state tensor in each neuron. This tensor gives the network encoded temporal context on the past inputs and outputs of the model, leading to potentially better prediction of models with temporal dependencies. Gated recurrent units (GRUs) \cite{bff0e6bd8f4a4f0d9735bf1728fb43ef} and long-short term memory (LSTM) \cite{Hochreiter1997} are two of the main architectures in this paradigm. One major limitation of such networks is the modeling of long temporal dependencies, due to the exploding and vanishing gradient problem \cite{pmlr-v139-rusch21a}. The transformer architecture has been extensively used due to its success overcoming these problems in the natural language processing community \cite{NIPS2017_3f5ee243}. This has encouraged its use in the modeling of optic-fiber channels \cite{Zhang2022, Zhu2023}. Our work in \cite{10382548} demonstrated the use of a variation of the transformer architecture, the convolutional attention transformer (CAT) \cite{10.5555/3454287.3454758} in the modeling of DMLs. Based on the structural similarities between neural ODEs and the residual connections in transformers \cite{li-etal-2022-ode}, we use the outputs of the DML rate equations to obtain a differentiable DML model. The model compared favourably to hyperparameter-optimized Volterra filters, TDNNs and LSTMs \cite{10382548}.

\section{End-to-end Optimization} \label{ch:e2e}
The throughput limitations of DMLs have also been studied from the DSP perspective. In this scope, we can distinguish 2 different trends: one of them aims to exploit the available bandwidth to increase spectral efficiency, while the other aims to compensate the DML-induced waveform distortion. Discrete multi-tone (DMT) \cite{8766826}, probabilistic constellation shaping (PCS) \cite{8627924} and GCS are the dominant technologies in the former category. DMT (the baseband, guided-system equivalent of orthogonal frequency-division multiplexing, OFDM) aims to exploit the available bandwidth through the use of multiple digital narrow-band sub-carriers, instead of using a unique one spanning all the available spectrum. In non-flat spectrum conditions, like the induced by PPR in DMLs, the SNR varies along different regions of the spectrum. The use of DMT allows higher granularity on the configuration for each sub-channel, optimizing it to the local conditions of each narrow band. PCS aims to replace uniform probability mass function (PMF) of the transmitted symbols by a different PMF in order to optimize a certain metric (energy per bit, SNR). GCS \cite{9852813} follows similar approach, but it modifies the energy allocated to each symbol instead of tweaking their probabilities. The combination of DMT and constellation shaping is usually called entropy loading \cite{8141854}, and it allows to adapt the symbol distribution to the spectral channel response, saving power in the high-SNR narrow bands while maximizing it in the low-SNR bands. Many DML implementations have included a combinations of these technologies to increase the throughput of DMLs \cite{Kim:23, 9186796, 9288898, Kottke:17}. Pulse shaping is another source of throughput optimization. Faster-than-Nyquist (FTN) \cite{5645721} signaling aims to exploit this by introducing a controlled amount of ISI in each symbol, correcting their introduced correlation on the RX side. This allows to increase the effective throughput, as long as the receiver is able to compensate both the TX- and the channel-induced ISI. \cite{9288898} combines FTN with entropy loading for increased spectral density of the DML bandwidth. Within distortion compensation schemes, most approaches are focused on either tailoring the peak-to-peak current to invert the nonlinear DML response or using pre-distortion and equalization techniques. \cite{9468374} provides a linearization method to correct the DML nonlinearities, while \cite{10070822} aims to suppress relaxation-oscillations using ML approaches. Even optical filters can be designed through semi-analytical approaches to find optimal cavity configurations \cite{10323176}. The development of DML-specific compensation approaches is also extensive, using different equalizer structures like Volterra \cite{photonics10101174, 8594656}, TDNN \cite{Reza:22}, RNN \cite{Xu:20} or deep belief networks \cite{Tian2020}.

Despite the remarkable performance leap obtained in some of the aforementioned data-driven approaches, their optimization is based on single-sided or sequential learning approaches: they do not optimize the TX and RX sides jointly. In sequential learning, each side of the communication link (TX and RX) is optimized keeping the configuration of the other one fixed, thus avoiding the need for gradient propagation across the transmission channel. E2E learning \cite{8054694} aims to condense several functions within TX (constellation shaping, pulse shaping, bit labelling) and RX (EQ, symbol decoding) into a single neural network, where the link configuration is optimized simultaneously at both ends based on a certain metric (loss function). Thus, the performance of the configuration obtained through E2E learning is limited mainly by the amount of DSP functions optimized and the accuracy of the modeling of the elements between TX and RX. This can pose a considerable challenge in the case of DML-based systems, where the building a differentiable model of the DML is not straightforward, as discussed in the previous section. Gradient-free approaches, based on derivative-free optimizers or reinforcement learning, have been proposed in the training of E2E \cite{8792076, 8645416, 10124361}. Nonetheless, they introduce severe computational overhead and their gradient approximations can lead to numerical optimization issues \cite{Yankov:22}. Several gradient-based have been proposed for DML (VCSEL) systems, implementing GCS \cite{Srinivasan:22} and pre-distortion + EQ \cite{Minelli:24} based on data-driven DML models. The present work aims to jointly optimize transmitter GCS and LPS and receiver EQ with the driving configuration of the DML ($I_{\mathrm{bias}}, I_{\mathrm{pp}}$), thus tailoring E2E learning to the specific characteristics of DML-based systems. This entails extending our DML modeling work in \cite{10382548}, obtaining a model that is able to predict the dynamics of the laser regardless of its biasing. 

\begin{figure}[t]
\includegraphics[width = \linewidth]{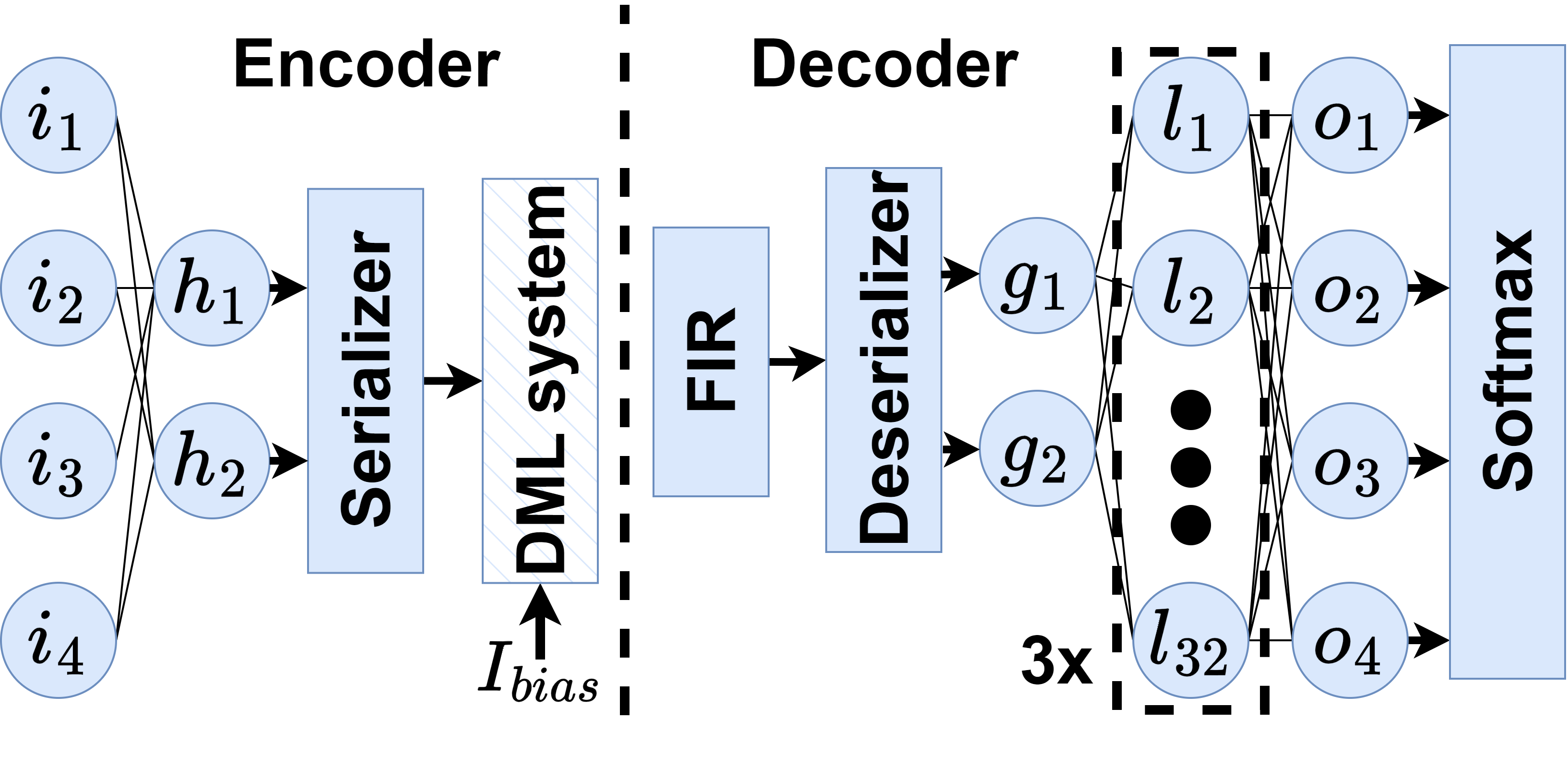}
\caption{Architecture of the proposed AE}
\label{fig:ae}
\end{figure}

The usual approach to E2E learning in communication systems is using autoencoders (AEs) \cite{Goodfellow-et-al-2016} as substitutes of the DSP pipeline \cite{8054694, 8433895, 8535453}. This is due to their functional resemblance to a communication link: they compress information subject to physical constraints (encoding) to then retrieve it with the minimum loss of information possible (decoding). The AE was created to compress and decompress information with minimal information loss, making it an ideal candidate for this task. It consists on an encoder, that performs the dimensionality reduction, and a decoder, that aims to retrieve the information compressed by the encoder (as shown in \cref{fig:ae}). AE-based optimization of communication systems \cite{8433895, 10124361} usually relies on an encoder to map an alphabet of one-hot encoded symbols $i_n$ to a vector of samples $h_n$ representing a symbol (assuming 2 sps in \cref{fig:ae}) to be transmitted. The decoder then maps the received samples $g_n$ back to scalars $o_n$ representing output probabilities, usually after a series of hidden layers, represented by $l_n$ in \cref{fig:ae}.  {It must be noted that the matrix multiplication in AEs introduces a significant complexity overhead to the TX and RX operation. Therefore, AEs are often used only to find optimal link configurations, that are then implemented in the form of less complex look-up tables and decision circuits. This allows to take advantage of the performance gains yielded by autoencoders while adjusting complexity to the requirements of optical transceivers \cite{10496171}.}  The standard AE \cite{Goodfellow-et-al-2016} is an FFNN, and therefore it lacks any memory mechanism in its structure. This explains the use of a finite impulse response (FIR) filter on the decoder side, that allows the AE to capture and compensate for ISI and undesired memory effects. The $o_n $ are then masked with a normalized exponential function (softmax), depicted in \cref{eq:softmax}.

\begin{equation}
    \sigma(o_n) = \frac{e^{o_n}}{\sum_{j=1}^N e^{o_j}}
    \label{eq:softmax}
\end{equation}

where $N$ is the number of neurons in the decoder output layer. The softmax function converts the single-dimensional output tensor in an array of normalized probabilities, thus giving insight on the certainty of the symbol prediction. As in any supervised learning scheme, the AE needs to be trained based on a certain loss function. Given that the goal of the link optimization is minimizing the loss of information over the transmission channel, a metric that captures this information loss is desirable. This leads to the use of categorical cross-entropy (CE) as an indirect estimator of mutual information between transmitted and received symbols \cite{10124361}. The expression of CE is shown in \cref{eq:ce}:

\begin{equation}
    J_{CE}(\theta) = \frac{1}{N} \sum_{n=1}^N \left[ - \sum_{m=1}^{M} i_m^{(n)} \log o_m^{(n)} (\theta) \right] \, ,
    \label{eq:ce}
\end{equation}

where $N$ represents the symbol batch size and $M$ represents the modulation order utilized. In the backpropagation stage of the training process, the gradient over trainable parameters $\theta$ is calculated to minimize CE. This leads to higher mutual information between both ends of the link, due to the approximation of its lower bound in \cref{eq:mi}:

\begin{equation}
    I(X;Y) \geq H(X) -  \hat{H}(X|Y) \, ,
    \label{eq:mi}
\end{equation}

where $H(X)$ is the entropy of the probability distribution of the transmitted symbols $X$ and $\hat{H}(X|Y)$ is the upper bound on the conditional entropy of $X$ given the probability distribution of the received symbols $Y$. $\hat{H}(X|Y)$ can be approximated by the CE between transmitted and received symbols, given that the expression of the true channel transition is unknown \cite{10124361}.

\begin{table}[t!]
\begin{tabular}{cccc}
\hline
\textbf{Parameter} & \textbf{Symbol} & \textbf{Value} & \textbf{Units} \\ \hline
Confinement factor & $\Gamma$ & 0.24 & - \\ \hline
Photon lifetime & $\tau_p$ & 2.60 & ps \\ \hline
Carrier lifetime & $\tau_c$ & 3.17 & ns \\ \hline
Transparent carr. dens. & $N_0$ & $2.00 \cdot 10^{24}$ & $m^{-3}$ \\ \hline
Active cross section & $\sigma_g$ & $3.34 \cdot 10^{-20}$ & $m^2$ \\ \hline
Eff. refractive index & $n_g$ & 4 & - \\ \hline
Active volume & V & $3.60 \cdot 10^{-17}$ & $m^3$ \\ \hline
Gain compression & $\epsilon$ & $2.00 \cdot 10^{23}$ & $m^3$ \\ \hline
Spont. emiss. factor & $\beta$ & $1.00 \cdot 10^{-3}$ & - \\ \hline
Diff. quant. eff. & $\eta_0$ & 0.20 & - \\ \hline
\end{tabular}
\centering
\caption{Utilized parameters for the rate equation laser simulation}
\label{tab:laseparams}
\end{table}

\begin{figure*}[t]
    \centering
    \includegraphics[width=\linewidth]{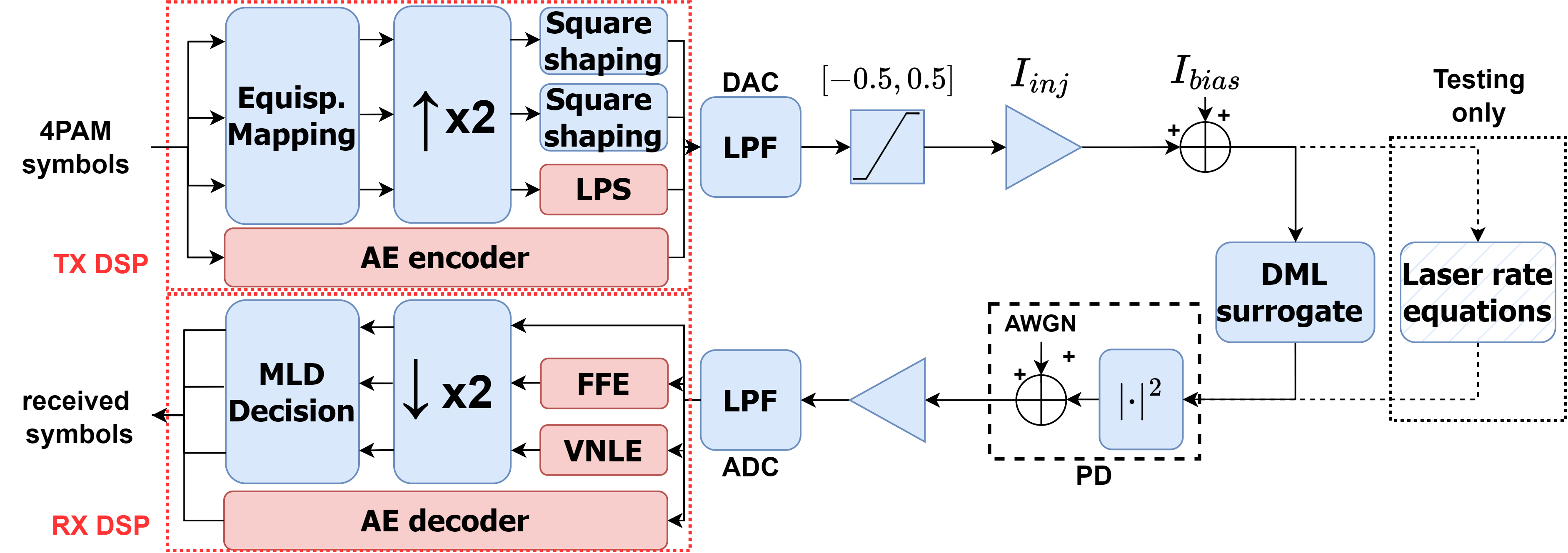}
    \caption{Block diagram of the proposed DML optimization approach. The trainable blocks are highlighted in red}
    \label{fig:setup}
\end{figure*}

\section{Simulation setup} \label{ch:setup}

The proposed DML system optimization is based on two main components: the laser surrogate model and the E2E optimization approach. The surrogate model aims solely to reproduce the laser behaviour as accurately as possible. The E2E approach finds the optimal system configurations subject to the constraints imposed by the surrogate model dynamics and its loss function, once the surrogate model has been trained. 

\subsection{Surrogate model}

\begin{figure}[t]
    \centering
    \includegraphics[width = \linewidth]{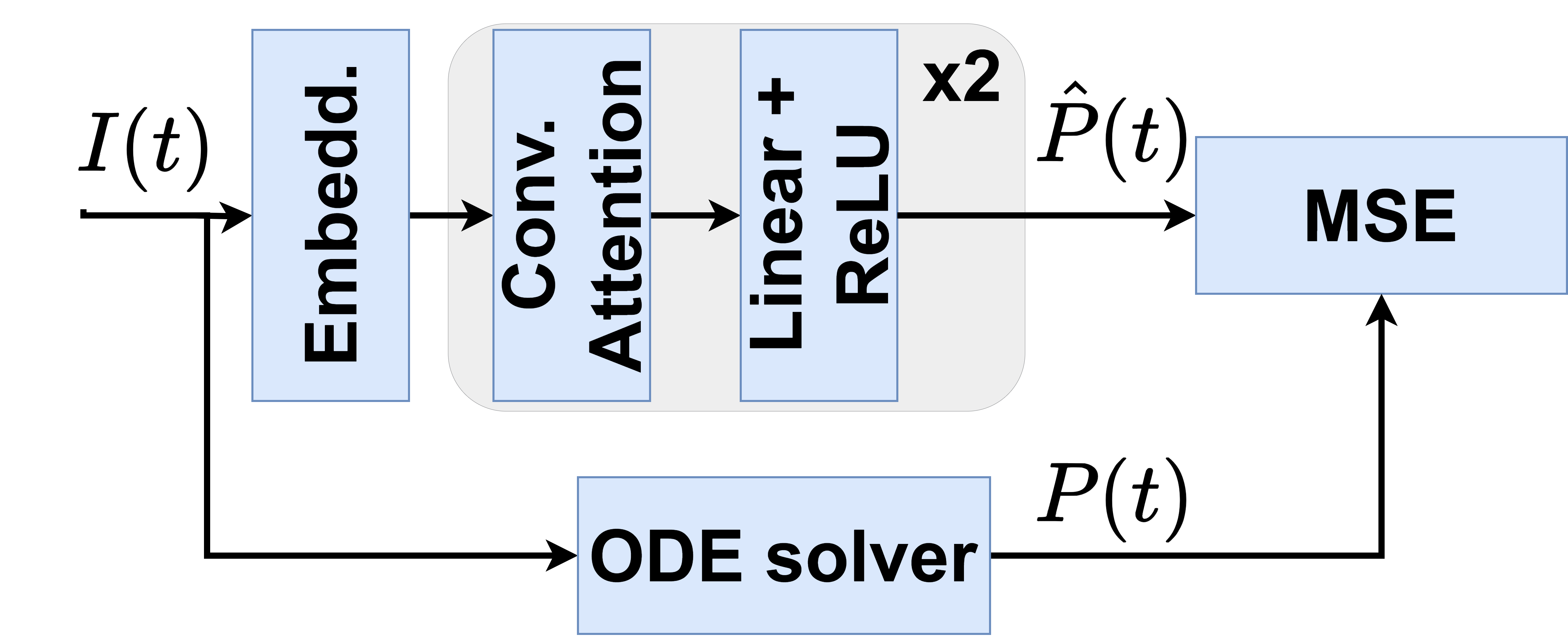}
    \caption{Architecture of the CAT. The network sub-blocks are described in Section~\ref{ch:setup}}
    \label{fig:convatt}
\end{figure}

The structure of the surrogate CAT model, depicted in \cref{fig:convatt}, is based on 3 building blocks: learned positional embeddings, convolutional attention and dense layers with ReLU activation. The positional embeddings aim to capture the position of each sample within the sequence, giving temporal context to the network. The convolutional attention block aims to discern the most relevant input samples for the calculation of each output sample through the use of matrix multiplication and learned convolutional filters. Lastly, the dense layers provide depth to the network, as they contain most of the trainable parameters within it. Although the model architecture, shown in \cref{fig:convatt}, is identical to \cite{10382548}, in this work we introduce a new data generation process. This allows to maintain model complexity with respect to our previous work while achieving a DML model agnostic to the waveform shape, the bias, and the peak-to-peak amplitude of the input sequence.

The surrogate modeling is performed in a data-driven fashion, using the laser rate equations (\cref{eq:phos,eq:carrs,eq:phase}) as source of data. The dataset of input waveforms to the rate equations is obtained by generating 4PAM-modulated symbols, pulse shaped at 2 samples per symbol (sps). The rate equation parameters utilized are specified in \cref{tab:laseparams}. The utilized parameters correspond to a  {generic DFB laser} with $f_R = 10.6$ GHz and a $f_{3\mathrm{dB}}$ of 25.5 GHz at $I_{\mathrm{bias}} = 75$ mA.  {Although the parameters are not chosen to match the specifications of a real device, they are of the same order of magnitude as those reported in the literature for single-mode quantum-well DFB lasers \cite{16895, 580827}.}  {It must be noted that $\phi(t)$ is not considered, given that the simulated setup is back-to-back. Therefore the values of $\alpha$ and $\kappa$ do not have an impact in the behaviour of the surrogate model.} The region of the modulation response beyond $f_R$ is of special interest due to the increased data throughput and significant waveform distortion introduced. The CAT model was therefore trained separately on 3 relatively high $R_s$, namely $\{15, 20, 25\}$ Gbaud. 

The generalizability in terms of waveform shaping is addressed by using 2 different types of pulse: square and stochastic pulses. The steep slope of the square pulse-shaped symbols allows to capture the transient dynamics of the DML response, while the stochastic pulses are used to prevent overfitting. The stochastic pulses are implemented through a 2-tap finite impulse response (FIR) filter, where the filter coefficients are drawn from a uniform distribution in the interval $[-0.5, 0.5]$. The pulses are normalized in amplitude to avoid distorting the 4PAM symbols. In order to maintain a high modelling accuracy regardless of the combination of $I_{\mathrm{bias}}$ and $I_{\mathrm{pp}}$ utilized, both quantities are randomized for the generation of the input waveform data set. This is achieved by randomizing $I_{\mathrm{bias}}$, drawing it from a uniform distribution in the interval $I_{\mathrm{bias}} = [50, 100]$ mA. $I_{\mathrm{pp}}$ is randomized in a similar fashion but constrained to the range $I_{\mathrm{pp}} = [0, 80]$ mA, in order to maintain a driving current significantly higher than the threshold current $I_{th} = 4.06$ mA for all $I_{\mathrm{bias}}$ and $I_{\mathrm{pp}}$ combinations. This will allow the optimization of the E2E learning approaches in a wide variety of different input currents, allowing the use of $I_{\mathrm{bias}}$ and $I_{\mathrm{pp}}$ as trainable parameters. 

Once the driving waveforms are generated, the sequences are then oversampled to 32 sps and low-pass filtered (LPF) to constrain their bandwidth to $0.9 R_s$. This is done to ensure the accuracy and convergence of the numerical solution to the rate equations, even if the surrogate model will operate on a 2-sps basis. The target (ground truth) output sequences for the CAT surrogate models are obtained by inputting the previously described driving waveforms to a numerical solver of the laser rate equations at a certain (randomized) $I_{\mathrm{bias}}$. The output photon density sequences from the rate equations are then downsampled and converted to optical power waveforms through \cref{eq:pout}. The loss function of the model is defined comparing the sequences from the rate equation solver to the predictions of the surrogate model using normalized root mean squared error (NRMSE). The use of NRMSE provides better loss interpretability, as the normalization applied makes the calculation of the loss relative to the amplitude of each generated sequence. Once the surrogate model is trained, its weights are fixed and it is used as part of an E2E optimization approach.

\subsection{E2E approach}

\begin{table}[t]
\begin{tabular}{cccc}
\hline
           & \textbf{BL/FFE}   & \textbf{VNLE}      & \textbf{AE}        \\ \hline
$I_{\mathrm{bias}}$ (mA) & 75       & 75        & Learnable \\ \hline
$I_{\mathrm{pp}}$ (mA)  & $[8,80]$ & $[8,80]$  & Learnable \\ \hline
Pulse shap.& Square   & Learnable & Learnable \\ \hline
GCS         & Equispaced  & Equispaced   & Learnable \\ \hline
Filt. taps & 0/21     & 272       & 11        \\ \hline
DML model  & Rate eqs. & CAT      & CAT       \\ \hline
Loss       & MSE      & MSE       & CE        \\ \hline
\end{tabular}
\caption{Parameters of the compensation approaches}%
\label{tab:params}
\end{table}

The E2E optimization approach aims to find the optimal combination of TX, RX and laser-driving parameters in order to minimize the probability of decision errors after detection.  {The approach should therefore reach optimal performance regardless of the targeted laser cavity structure, as long as the surrogate DML model is able to reproduce the laser behavior accurately. Thus, our approach paves the way for the E2E optimization of optical communication system based on external modulators, like electro-absorption modulators (EAMs), as long as data availability allows for the training of a sufficiently accurate surrogate model. }
The investigated back-to-back IM/DD system is represented in \cref{fig:setup}. The modulation is based on equiprobable 4PAM symbols, upsampled to 2 sps. The TX establishes the GCS constellation intensity levels and the 2-tap pulse shaping in order to generate the modulation current sequences. The hardware limitations of the digital-to-analog and analog-to-digital converters (DAC/ADC) are modeled as FIR LPFs, with a bandwidth $B_{LPF} = 0.9 \cdot R_s$.  {The impact of impairments associated with DAC and ADC is not considered in this paper. However, the AE is trained to take into account the limited time resolution and BW of DAC and ADC.} After filtering, the sequences are constrained to the range $[-0.5, 0.5]$ and then amplified to $[-40, 40]$ mA for a maximum peak-to-peak $I_{\mathrm{pp}}$ of 80 mA, in order to match the surrogate data generation. $I_{\mathrm{bias}}$ is again constrained to the range $[50, 100]$ mA. During training, the emulation of the DML response is carried out by the surrogate model, as its architecture allows for automatic differentiation. The testing is however conducted on the laser rate equations, using the same parameters as for data generation. This allows to obtained more reliable performance metrics, as the surrogate may distort them due to modeling inaccuracies.

\begin{figure}[t]
    \centering
    \includegraphics[width = \linewidth]{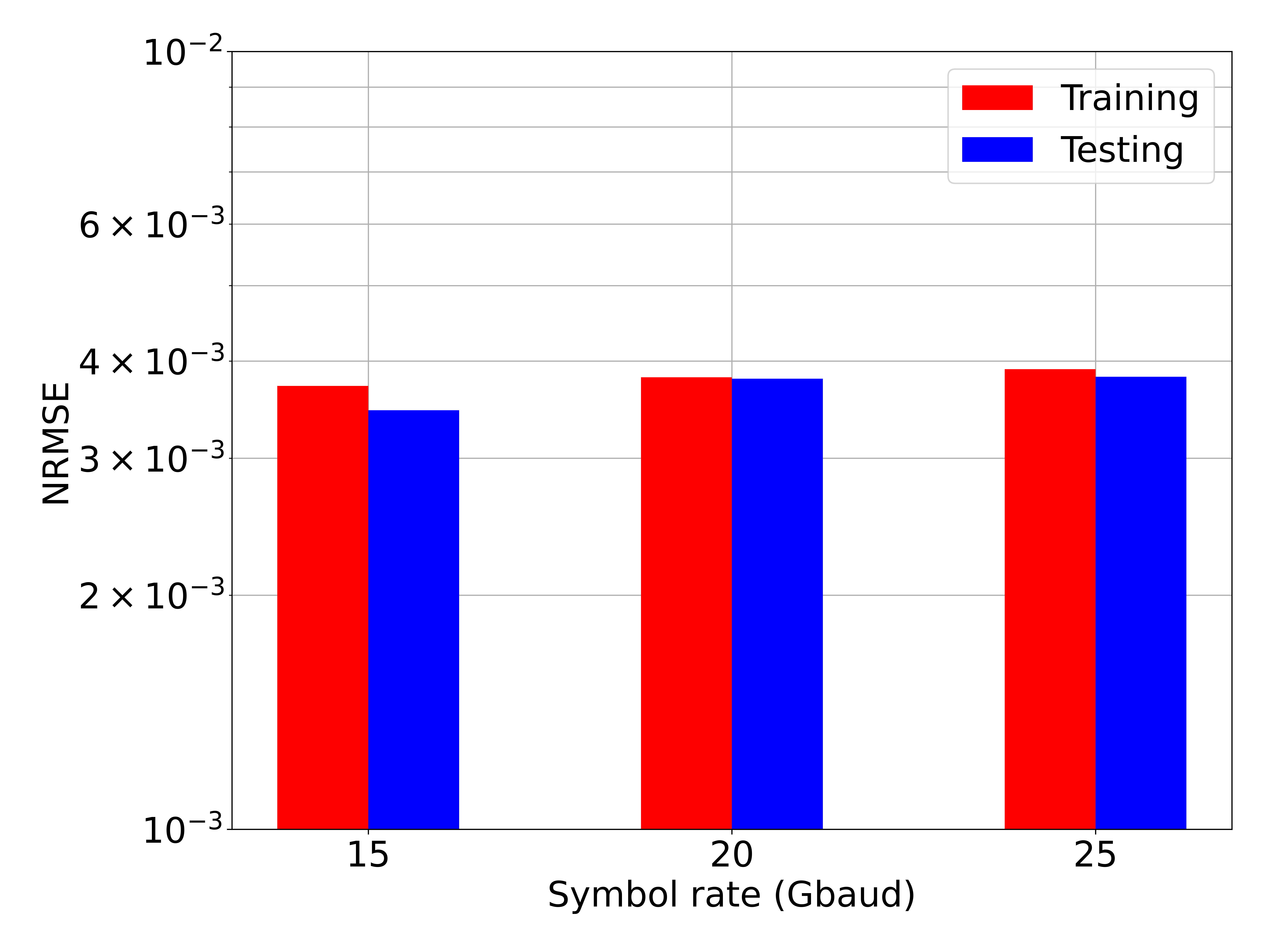}
    \caption{Training and testing NRMSE scores of the CAT surrogate, using the laser rate equations as ground truth}
    \label{fig:surr_res}
\end{figure}

\begin{figure*}[!ht]
    \centering
    \includegraphics[width= \linewidth]{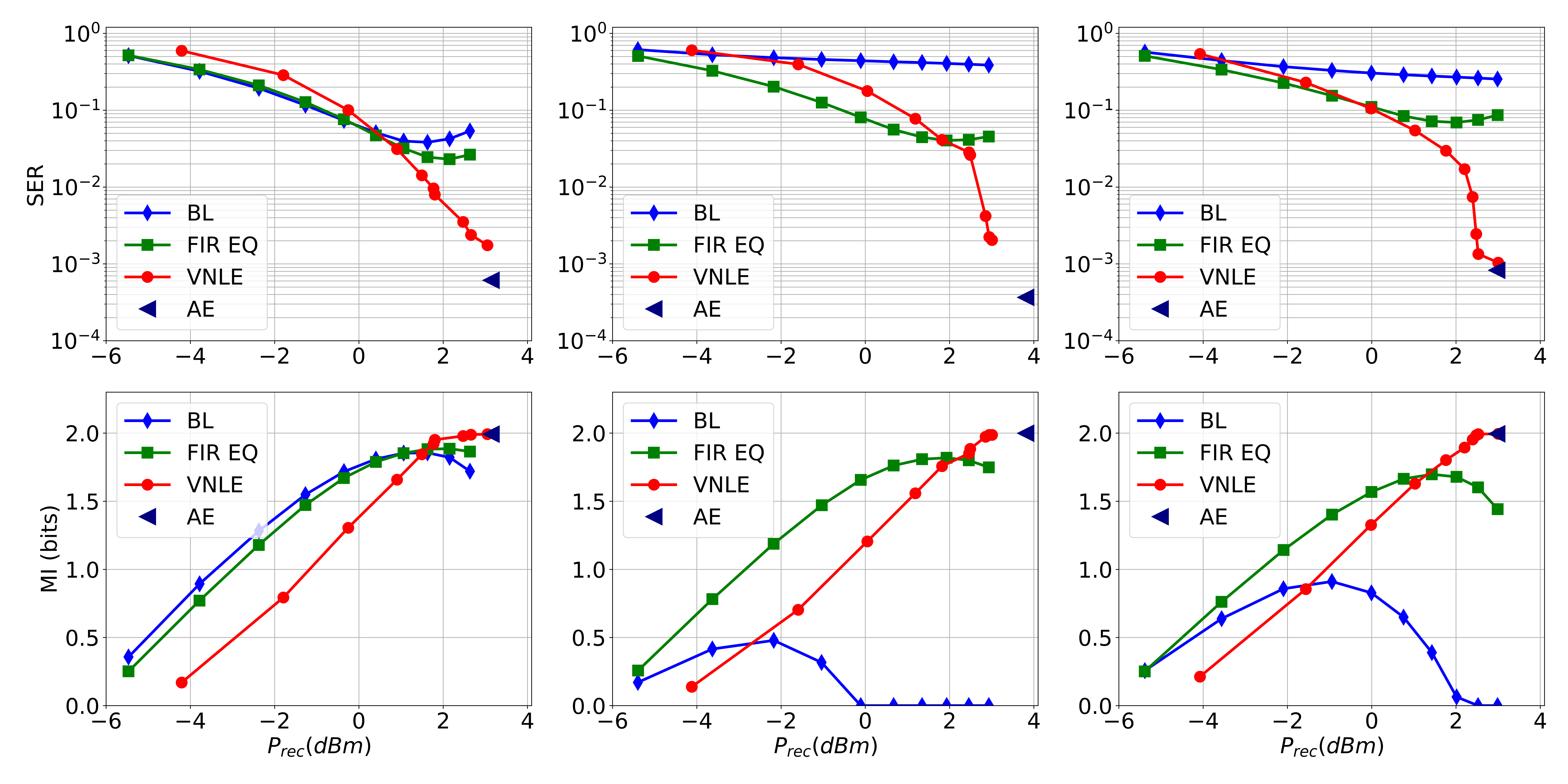}
    \caption{ {Symbol error rate (SER, top) and mutual information (MI, bottom) results at symbol rates $R_{s}=\{15, 20, 25\}$ Gbaud (left, middle, right) for the uncompensated baseline (BL), finite impulse response equalizer (FIR EQ), Volterra nonlinear equalizer (VNLE) and autoencoder (AE) setups}}
    \label{fig:results}
\end{figure*} 

After generating the output sequences from the DML models, additive white Gaussian noise (AWGN) samples are added to the signal after square-law detection. The noise variance is fixed to yield 22 dB electrical signal-to-noise ratio (SNR) at the highest average $P_\mathrm{rec}$ (corresponding to $I_{\mathrm{pp}} = 80$ mA). The value of $I_{\mathrm{bias}}$ does not impact $P_\mathrm{rec}$, as only the peak-to-peak power is considered in it.  Lastly, the adaptive RX DSP performs 2-sps-wise EQ, downsampling and symbol decision. The SER and MI performance metrics are obtained by comparing the estimated symbol probabilities at RX with the originally sent sequence at TX, using maximum likelihood detection (MLD) or softmax activation depending on the approach. 
The compensation of the system is based on four different approaches: the baseline uncompensated system (BL), an RX-side FIR FFE, a second-order VNLE with a transmitter LPS and an AE. Table~\ref{tab:params} lists the specific parameters for each approach. The FFE is meant to emulate the conventional approach to DML-based system optimization, while the VNLE and AE give insight on the performance advantage of E2E models. It must be noted that only the AE handles $I_{\mathrm{pp}}$ and $I_{\mathrm{bias}}$ as trainable parameters, while all the other models are kept at fixed bias and swept through various $I_{\mathrm{pp}}$ (and therefore $P_{\mathrm{rec}}$) values. 
Fig.~\ref{fig:ae} shows the encoder and decoder AE architecture from input symbol $i_n$ to output symbol likelihoods $o_n$. The encoder, or TX side, is based on a single linear layer to map the input one-hot encoded vectors into 2-sample pulses. All the obtained pulses are then serialized into 1024-sample time sequences to be used as driving current to the DML. The $I_{\mathrm{bias}}$ level is added to the signal after filtering and amplification, as explained previously. On the decoder (RX) side, the received sequences are first FIR filtered in order to provide a memory mechanism to the AE. After deserialization, 3 leaky-ReLU-activated feedforward layers with softmax activation at the output convert the filtered samples into symbol probabilities for decision. The AE loss is then calculated using cross-entropy (CE) between the originally transmitted symbols and their assigned symbol probability at the receiver.

\section{Results} \label{ch:results}
\subsection{Surrogate model}
The surrogate model training compromises $2^{23}$ samples in sequences of 1024 samples, while the testing data set includes $2^{17}$ samples. The randomization of pulse shape and laser driving configurations is performed on a per-sequence basis. The training and testing NRMSE scores obtained by the CAT model are depicted in \cref{fig:surr_res}. Throughout the three analyzed $R_s$, the testing NRMSE loss is slightly lower than its training counterpart, giving no sign of overfitting. Another interesting trend is the slight increase in NRMSE as $R_s$ increases. This trend might be related to the predominance of the nonlinear effects at higher $R_s$, leading to potentially more complex DML dynamics. In any case, the three models perform well below the $1\%$ mark.

\subsection{E2E approach}
The symbol dataset of the E2E approaches compromises $2^{20}$ symbols, with an 80/20 partition between training and validation, respectively. The dataset is split in mini-batches of 512 symbols for exploiting parallelization to reduce the training time. The $I_{\mathrm{pp}}$ to all the models except the AE is swept in the range $[8, 80]$ with a step of 8 mA between consecutive levels.

Even though all the approaches are subject to the same $I_{\mathrm{pp}}$ constraints, the E2E approaches are able to exploit the DML transient response through LPS (and GCS in the case of the AE), yielding higher $P_\mathrm{rec}$. The AE is able to optimize its $I_{\mathrm{pp}}$ dynamically within the constrained range, therefore only one AE power level was analyzed. The training is iterated over three different symbol rates: 15, 20 and 25 Gbaud, matching the used in the surrogate training. The MI and SER results tested on the laser rate equations are shown as a function of $P_\mathrm{rec}$ in Fig.~\ref{fig:results}. Based on the figure, the AE delivers the best SER and MI performance overall, hinting that the optimization of $I_{\mathrm{bias}}$ level could have a higher impact on performance than the equalization in certain cases. This is more accentuated on the lower $R_s$ than the higher ones, where the waveform distortion worsens the SNR, giving equalization a higher relative impact. Another interesting metric is the optimal $I_{\mathrm{bias}}$ obtained by the AE, resulting in $\{62.31, 69.31, 70.30\}$ mA at $\{15, 20, 25\}$ Gbaud, respectively. This trend provides interesting insight on the relation between the $R_s$ and the optimal $I_{\mathrm{bias}}$ conditions. For the $R_s$ analyzed, the E2E approaches show a clear performance advantage over the BL and RX-only optimization. This is expected, given that the lack of nonlinearity in the latter makes the complete compensation of the DML-induced distortion infeasible at high symbol rates. 
The better performance of the E2E approaches serves as a further validation of the surrogate, as a poor model of the DML could lead to low performance when tested on the rate equations.

\section{Conclusion} \label{ch:conclusion}
We propose a novel end-to-end, directly modulated laser optimization approach using a differentiable data-driven model as laser surrogate, allowing the propagation of gradients between transmitter and receiver. The surrogate is built based on the laser rate equations using various bias and peak-to-peak current values, in order to make it robust to such values. We compare different system architectures with conventional receiver-side optimization, varying the received optical power and the symbol rate of the simulation. The proposed autoencoder approach including bias and peak-to-peak current optimization shows significant performance gain compared to its receiver-side counterpart, showcasing the potential of end-to-end approaches in the optimization of directly modulated laser systems. 

\section*{Funding} The Villum Fonden (VI-POPCOM VIL54486) and Villum YIP OPTIC-AI (no. VIL29334) projects are acknowledged.


\bibliography{refs}


\end{document}